\documentclass[superscriptaddress,aps,prd,nofootinbib,reprint,onecolumn]{revtex4-2}
\pdfoutput=1

\usepackage{graphicx}
\usepackage[colorlinks,allcolors=blue]{hyperref} 
\usepackage{amsmath,amssymb,bm}
\usepackage[T1]{fontenc}
\usepackage{mathtools}

\begin{document}

\hfill TU-1267

\title{
Resolving the QCD Axion Domain Wall Problem with a Light Axion
}

\author{Junseok Lee}
\email{lee.junseok.p4@dc.tohoku.ac.jp}
\affiliation{Department of Physics, Tohoku University, Sendai, Miyagi 980-8578, Japan}
\author{Kai Murai}
\email{kai.murai.e2@tohoku.ac.jp}
\affiliation{Department of Physics, Tohoku University, Sendai, Miyagi 980-8578, Japan}
\author{Fuminobu Takahashi}
\email{fumi@tohoku.ac.jp}
\affiliation{Department of Physics, Tohoku University, Sendai, Miyagi 980-8578, Japan} 
\affiliation{Kavli IPMU (WPI), UTIAS, University of Tokyo, Kashiwa 277-8583, Japan}
\author{Wen Yin}
\email{wen@tmu.ac.jp}
\affiliation{Department of Physics, Tokyo Metropolitan University, Tokyo 192-0397, Japan}

\begin{abstract}
We propose two novel solutions to the domain wall problem of the QCD axion by introducing a massless or 
light axion that also couples to gluons.
The first solution applies when the new axion forms strings after inflation. Due to its mixing with the QCD axion, domain walls of the QCD axion are bounded by these strings and confined into cosmologically safe string bundles. This scenario predicts the existence of such string bundles, which may survive until today and leave observable signatures, such as gravitational waves, cosmic birefringence, and CMB anisotropies.
The simultaneous detection of the QCD axion and any of these cosmological signatures would serve as a smoking-gun signal.
The second solution assumes a homogeneous initial condition for the new axion. If it is sufficiently light, its potential temporarily induces a bias in the QCD axion potential before the onset of oscillations, rendering the domain walls unstable. In both scenarios, the Peccei–Quinn mechanism remains effective, and the strong CP problem is not reintroduced.
We identify the viable parameter regions and discuss the resulting dark matter abundance.
\end{abstract}

\maketitle

\section{Introduction}
\label{sec: intro}

The Peccei–Quinn (PQ) mechanism~\cite{Peccei:1977hh,Peccei:1977ur} provides a compelling solution to the strong CP problem by promoting the $\theta$ parameter of QCD to a dynamical field, known as the QCD axion~\cite{Weinberg:1977ma,Wilczek:1977pj}.
The QCD axion is a pseudo Nambu–Goldstone (NG) boson associated with the spontaneous symmetry breaking (SSB) of the global $\mathrm{U}(1)_{\mathrm{PQ}}$ symmetry.  
In addition to solving the strong CP problem, it is also a well-motivated dark matter candidate due to its weak interactions and long lifetime.
See, e.g., Refs.~\cite{Kim:2008hd,Arias:2012az,Marsh:2015xka,DiLuzio:2020wdo,OHare:2024nmr} for reviews.  

The cosmological production of the QCD axion depends on whether the $\mathrm{U}(1)_{\mathrm{PQ}}$ symmetry is spontaneously broken before or after the end of inflation, referred to as the pre-inflationary and post-inflationary scenarios, respectively.
The post-inflationary scenario typically suffers from the domain wall problem~\cite{Zeldovich:1974uw}. After the SSB of the $\mathrm{U}(1)_\mathrm{PQ}$ symmetry, cosmic strings are formed. When the Universe cools down to around the QCD scale, the QCD axion acquires a potential and starts oscillating around the nearest minimum. This leads to the formation of domain walls attached to the strings. 

The fate of the string-wall network crucially depends on the domain wall number $N_{\rm DW}$, defined as the number of domain walls attached to a single string.
If $N_\mathrm{DW} = 1$, the string-wall network collapses due to the wall tension and disappears shortly after its formation.
Then, the energy stored in the network is released as axion fluctuations, which can make up the dark matter~\cite{Davis:1986xc} (see also Refs.~\cite{Redi:2022llj,Gonzalez:2022mcx,Aldabergenov:2024fws} for a longer lifetime of domain walls due to inflationary dynamics).
If $N_{\rm DW} > 1$, on the other hand, the network becomes stable and can dominate the energy density of the Universe, leading to the cosmological domain wall problem~\cite{Zeldovich:1974uw}. For instance, in the DFSZ axion model~\cite{Dine:1981rt,Zhitnitsky:1980tq}, it is known that $N_{\rm DW}$ is either 3 or 6.
In the KSVZ axion model, $N_{\rm DW}$ is determined by the number of heavy PQ quarks; however, there is no strong theoretical reason that enforces $N_{\rm DW} = 1$ in general.
Addressing the domain wall problem in the post-inflationary scenario is therefore essential for constructing a viable axion cosmology.

Various solutions have been proposed to resolve the domain wall problem of the QCD axion. One is to introduce an explicit PQ symmetry breaking term that generates a potential bias between different vacua, making the domain walls unstable and causing the network to collapse~\cite{Sikivie:1982qv,Hiramatsu:2012sc}. However, this approach could spoil the PQ mechanism unless the CP phase of the explicit breaking is carefully chosen.
See also Refs.~\cite{Barr:2014vva,Takahashi:2015waa,Reig:2019vqh,Caputo:2019wsd} for cases where the PQ symmetry is dynamically broken.
Another approach is the Lazarides–Shafi (LS) mechanism~\cite{Lazarides:1982tw,Sato:2018nqy,Chatterjee:2019rch}, which identifies the degenerate vacua through gauge or global transformations, effectively reducing the domain wall number to unity. In this case, the axion strings decompose into $N_{\rm DW}$ (or more) hybrid or Alice strings, each attached to (at most) a single domain wall, thus solving the cosmological domain wall problem, because such a string-wall network collapses soon due to the wall tension.
Nonetheless, realistic model-building along these lines, especially in the DFSZ framework, tends to require rather elaborate extensions.

In addition to the QCD axion, many other axions with masses spanning a wide range may exist, as suggested by the axiverse scenario~\cite{Arvanitaki:2009fg}. Such axions, often referred to as axion-like particles (ALPs), are predicted in various extensions of the Standard Model, particularly in string theory~\cite{Witten:1984dg,Conlon:2006tq,Svrcek:2006yi,Arvanitaki:2009fg,Cicoli:2012sz}. These axions generically mix with one another, giving rise to a variety of cosmological phenomena. In particular, the mixing between the QCD axion and an ALP can lead to rich dynamics such as level crossing~\cite{Kitajima:2014xla,Daido:2015bva,Daido:2015cba,Ho:2018qur,Cyncynates:2023esj,Murai:2024nsp}, modifications to the misalignment mechanism~\cite{Murai:2023xjn}, decaying dark matter~\cite{Higaki:2014qua}, the clockwork mechanism~\cite{Higaki:2015jag,Higaki:2016jjh,Higaki:2016yqk,Long:2018nsl}, and inflationary dynamics~\cite{Takahashi:2019qmh,Kobayashi:2019eyg,Takahashi:2019pqf,Kobayashi:2020ryx,Narita:2023naj}.

Recently, we investigated the dynamics of topological defects in the mixed axion system~\cite{Lee:2024xjb,Lee:2024toz}, where various novel phenomena emerge compared to the single axion system.
One is the formation of string bundles~\cite{Higaki:2016jjh, Long:2018nsl}, and the other is a transient bias~\cite{Lee:2024toz}, both of which play important roles in the dynamics of the axionic string-wall network. A string bundle refers to an isolated object in which strings associated with multiple axions are bound together by domain walls. It effectively corresponds to the cosmic string of a light axion at low energies. The transient bias arises when the potential of a light axion temporarily acts as a bias on the string-wall network before the onset of its oscillations.

In this paper, we propose two novel solutions to the domain wall problem of the QCD axion by introducing a massless or light ALP coupled to gluons. The first solution applies when the ALP follows a post-inflationary scenario. If the domain wall number of the ALP is unity, string bundles composed of QCD axion and ALP strings are readily formed, replacing the original domain wall network with a cosmologically safe string network. The second solution applies when the ALP follows a pre-inflationary scenario. In this case, a transient bias arising from the ALP potential destabilizes the domain wall network and induces its collapse. Here, the ALP remains frozen in the early Universe and effectively acts as an explicit breaking of the PQ symmetry. However, since the axions are eventually driven to the potential minimum, the strong CP phase vanishes in the present Universe, and the strong CP problem is not reintroduced. In both scenarios, a certain amount of QCD axion is produced during the network evolution, which can contribute to dark matter.

The rest of this paper is organized as follows.
In Sec.~\ref{sec: DW problem}, we briefly review the domain wall problem of the QCD axion and explain the possible solutions.
We describe the solution by string bundles and discuss the dark matter production in Sec.~\ref{sec: string bundle}.
In Sec.~\ref{sec: transient}, we describe the other solution by transient bias.
Section~\ref{sec: summary} is devoted to the summary and discussion of our results.

\section{Domain wall problem of the QCD axion}
\label{sec: DW problem}

Throughout this paper, we assume the post-inflationary scenario for the QCD axion so that cosmic strings are formed after the SSB of the $\mathrm{U}(1)_{\mathrm{PQ}}$ symmetry. 
Let $S$ be the PQ scalar that hosts the QCD axion $a$:
\begin{align}
    S = \frac{f_a}{\sqrt{2}}e^{i \frac{a}{f_a}},
\end{align}
where $f_a$ is the decay constant of the QCD axion, and the radial degrees of freedom are omitted for simplicity. For our purpose it is sufficient to adopt this minimal model with a single PQ scalar. See Ref.~\cite{Lee:2024toz} for the case of multiple PQ scalars.

The QCD axion $a$ has an anomalous coupling to gluons and acquires a potential from non-perturbative effects of QCD as
\begin{align}
    V_\mathrm{QCD}(a)
    =
    \chi(T) \left[ 
        1 - \cos \left( N_{\rm DW}\frac{a}{f_a} \right)
    \right]
    \ ,
\end{align}
where  $N_{\rm DW}$ is an integer called the domain wall number. 
Note that, in the axion literature, the decay constant is conventionally defined as $F_a \equiv f_a/N_{\rm DW}$ rather than $f_a$.
Here, we define the origin of $a$ so that $V_\mathrm{QCD}$ takes its minimum value at $a = 0$.

At high temperatures, the topological susceptibility, $\chi(T)$, is suppressed.
Its temperature dependence is represented by 
\begin{align}
    \chi(T)
    \simeq
    \left\{
        \begin{array}{ll}
            \chi_0 & \quad (T < T_\mathrm{QCD})
            \\
            \chi_0 \left( \dfrac{T}{T_\mathrm{QCD}} \right)^{-n} & \quad (T \geq T_\mathrm{QCD})
        \end{array}
    \right.
    \ ,
\end{align} 
where we adopt $\chi_0 \simeq (75.6\,\mathrm{MeV})^4$, $T_\mathrm{QCD} \simeq 153\,\mathrm{MeV}$, and $n \simeq 8.16$~\cite{Borsanyi:2016ksw}. 
For later convenience, we define the temperature-dependent mass and the constant mass at low temperatures,
\begin{align}
    m_a(T)
    \equiv 
    \frac{N_{\rm DW}\sqrt{\chi(T)}}{f_a}
    \ , \quad 
    m_{a0} 
    \equiv 
    m_a(T<T_\mathrm{QCD})
    \ .
\end{align}

Once $m_a(T)$ or $m_{a0}$ becomes comparable to the Hubble parameter, the axion starts to roll down to the nearest potential minimum, and $N_{\rm DW}$ domain walls appear, attached to each string.
If $N_{\rm DW} = 1$, the network of the strings and walls collapses due to the wall tension and rapidly disappears.
On the other hand, if $N_{\rm DW} > 1$, the string-wall network is stable and approaches the so-called scaling regime, where each Hubble volume contains $\mathcal{O}(1)$ walls.
In this regime, the energy density of domain walls decreases more slowly than matter and radiation, and eventually dominates the Universe.
From the observed temperature anisotropies of the CMB, the domain wall tension is constrained as~\cite{Zeldovich:1974uw,Vilenkin:1984ib}
\begin{align}
\label{eq:const_on_sigma}
    \sigma 
    \lesssim
    (1\,\mathrm{MeV})^3
    \ .
\end{align}
Since the domain wall tension is given by $\sigma_a  \simeq  8 m_a f_a^2/N_{\rm DW}^2$, which is larger than the QCD scale, this condition cannot be satisfied for the QCD axion.
Thus, the post-inflationary scenario with $N_{\rm DW} > 1$ suffers from the cosmological domain wall problem~\cite{Zeldovich:1974uw}.

One possible solution to the domain wall problem is to introduce a potential bias.
If the QCD axion potential receives an additional contribution that lifts the degeneracy among the $N_{\rm DW}$ vacua of $V_\mathrm{QCD}$, the resulting pressure difference across the domain walls drives the decay of the string-wall network.
However, unless the minima of $V_\mathrm{QCD}$ and those of the bias term are finely aligned, the true vacuum is generically displaced from the CP-conserving minimum of $V_\mathrm{QCD}$.
In such cases, the strong CP phase remains nonzero, and the strong CP problem reappears in the form of a fine-tuning problem in the phase of the bias term~\cite{Barr:1992qq,Kamionkowski:1992mf,Holman:1992us,Kawasaki:2014sqa,Chang:2023rll,Beyer:2022ywc}.

In the following sections, we introduce another light axion coupled to gluons to solve the domain wall problem.

\section{Solution with string bundles}
\label{sec: string bundle}

Here we introduce an additional axion or ALP, $\phi$, which also couples to gluons.
Let the corresponding PQ scalar $\Phi$ be written as
\begin{align}
    \Phi = \frac{f_\phi}{\sqrt{2}} e^{i \frac{\phi}{f_\phi}},
\end{align}
where $f_\phi$ denotes the decay constant of $\phi$, and the radial mode is omitted for simplicity.
In this section, we assume the post-inflationary scenario for both $a$ and $\phi$.

As in the KSVZ axion model~\cite{Kim:1979if,Shifman:1979if}, 
one can couple $\phi$ to gluons by introducing a heavy, vector-like PQ quark.
Non-perturbative QCD effects then generate a potential 
\begin{align}
    V_\mathrm{QCD}(a,\phi)
    =
    \chi(T) \left[ 
        1 - \cos \left( N_{\rm DW}\frac{a}{f_a} + \frac{\phi}{f_\phi} \right)
    \right]
    \ .
\end{align}
We assume that the domain wall number of $\phi$ is equal to unity; this property is essential to our resolution of the QCD-axion domain-wall problem.

With this potential, the combination of $a$ and $\phi$ in $V_\mathrm{QCD}$ plays the role of the QCD axion, while the perpendicular combination in this basis remains massless.
The effective QCD axion $A$ and the massless axion $A_L$ are given by 
\begin{align}
    A & \equiv  F \left( N_{\rm DW}\frac{a}{f_a} + \frac{\phi}{f_\phi} \right) ,
    \\
    A_L & \equiv  F \left( -\frac{a}{f_\phi} + N_{\rm DW}\frac{\phi}{f_a} \right)
    \end{align}
with
\begin{align}
    F \equiv  \frac{f_a f_\phi}{\sqrt{N_{\rm DW}^2 f_\phi^2 + f_a^2}}.
\end{align}
The mass of $A$ is given by
\begin{align}
    m_A(T)
    \equiv
    \frac{\sqrt{\chi(T)}}{F}
    \ .
\end{align}
For later convenience, we define $m_{A0} \equiv m_A(T < T_\mathrm{QCD})$.

When $S$ and $\Phi$ acquire nonzero vacuum expectation values, two types of strings—$a$-strings and $\phi$-strings—are formed. Later, when the axion mass becomes comparable to the Hubble parameter, $m_A \sim H$, domain walls appear. These walls can be bounded by both types of strings. More specifically, each $a$-string is attached to $N_{\rm DW}$ domain walls, while each $\phi$-string is attached to a single wall.

Such a string-wall network subsequently transforms into string bundles consisting of a single $a$-string and $N_{\rm DW}$ $\phi$-anti-strings (or equivalently, a single $a$-anti-string and $N_{\rm DW}$ $\phi$-strings)~\cite{Higaki:2016jjh,Long:2018nsl,Eto:2023aqr,Lee:2024toz}.
We show the sketch of string bundle formation in Fig.~\ref{fig: bundle} (see also Fig.~\ref{fig: string-wall snapshot} in Appendix \ref{appendix B}). It is crucial that the domain wall number of $\phi$ is unity; if it were greater than one, the string-wall network would persist instead of collapsing into string bundles~\cite{Lee:2024toz}.
\begin{figure}[t]
    \centering
    \includegraphics[width=.4\textwidth]{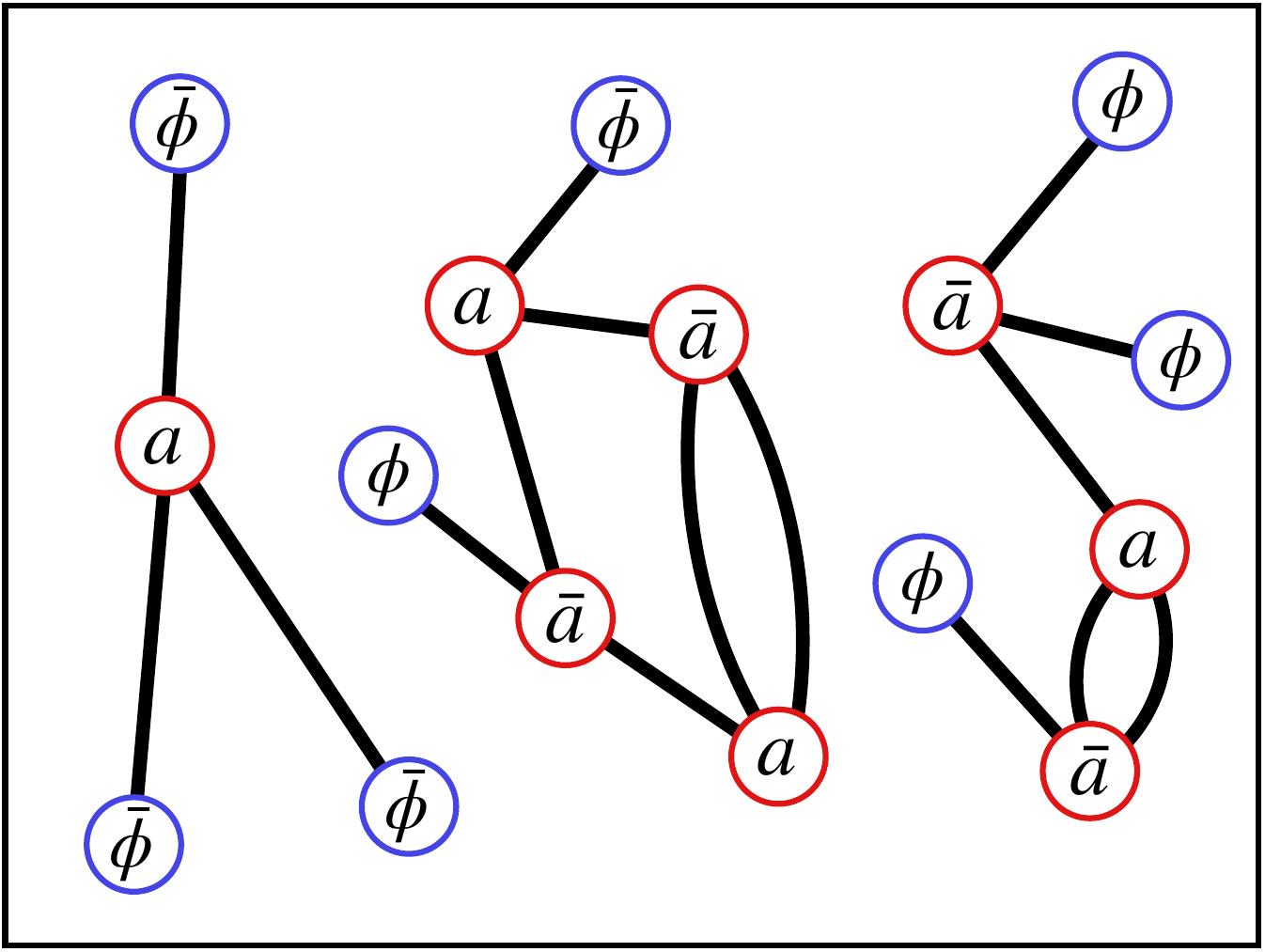}
    \hspace{5mm}
    \includegraphics[width=.4\textwidth]{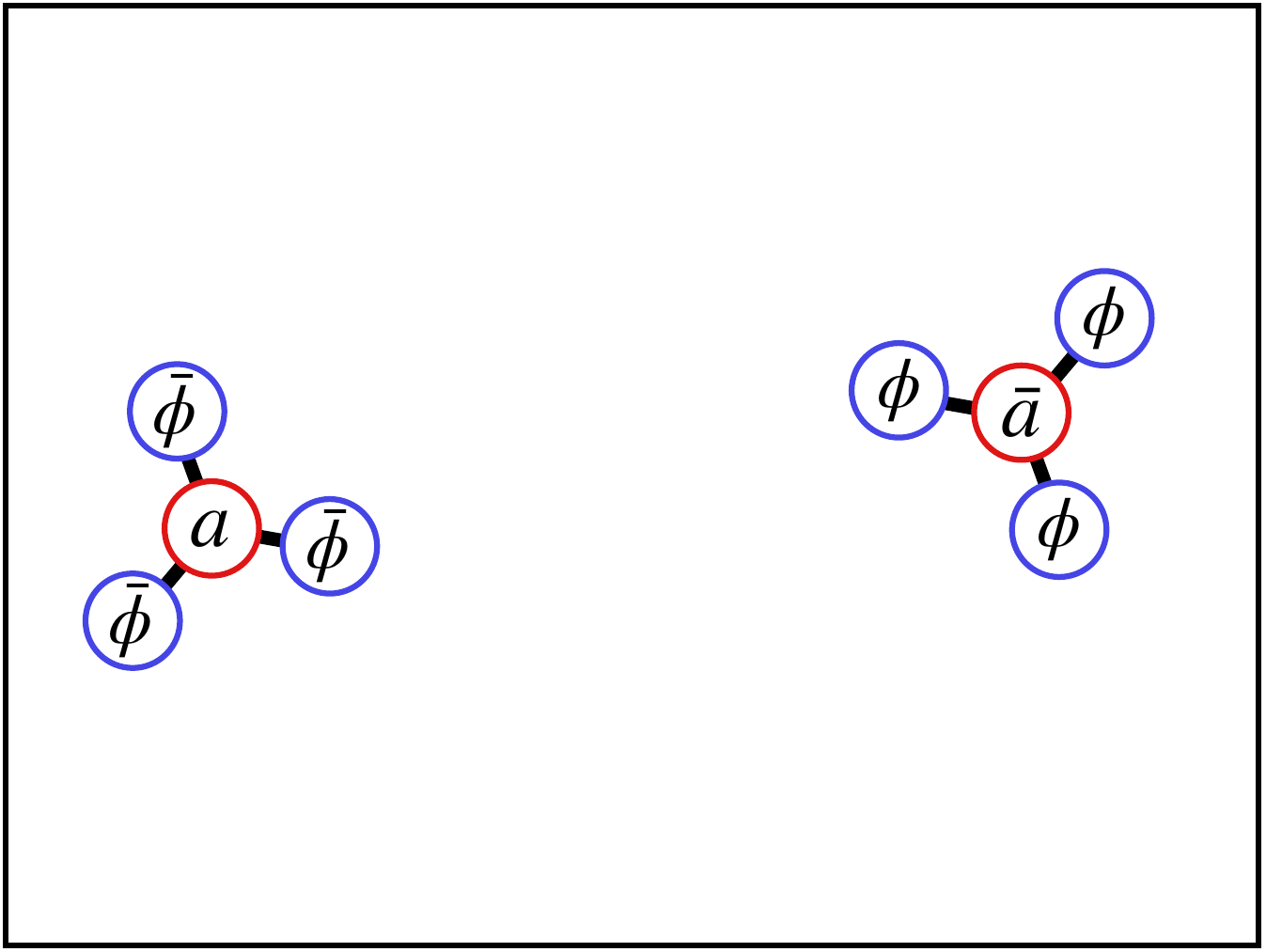}
    \caption{%
       Schematic illustration of string bundle formation.
       The circles with $a$ and $\phi$ represent cosmic strings of $a$ and $\phi$, respectively, and the circles with $\bar{a}$ and $\bar{\phi}$ represent their corresponding anti-strings.
       The black lines connecting strings represent domain walls.
       Left panel:
       When the domain wall tension is too small to pull the strings together, the string-wall network stretches over cosmological scales.
       Right panel:
       Once the domain walls pull the strings together, the strings and their anti-strings annihilate with each other, and a specific combination of strings forms a string bundle.
    }
    \label{fig: bundle}
\end{figure}

After the network has fully transitioned into string bundles,  the domain walls are confined within the string bundles, and their energy density scales like that of cosmic strings.
The resulting string bundle can be interpreted as a cosmic string associated with the massless axion $A_L$. As a result, only cosmic strings remain in the late Universe, without dominating the energy density. 
Interestingly, when the tension satisfies $\mu_\phi < (10^{14\text{--}15}\,\mathrm{GeV})^2$~\cite{Lopez-Eiguren:2017dmc,Chang:2021afa,Servant:2023mwt}, current observations place no strong constraints.
Thus, the domain wall problem is resolved.\footnote{
The same dynamics were pointed out in Ref.~\cite{Takahashi:2020tqv}  in the context of connecting the cosmic birefringence to the QCD axion.
}We will later discuss the preferred value of the tension that leads to the correct abundance of the QCD axion dark matter.

\begin{figure}[t]
    \centering
    \includegraphics[width=.8\textwidth]{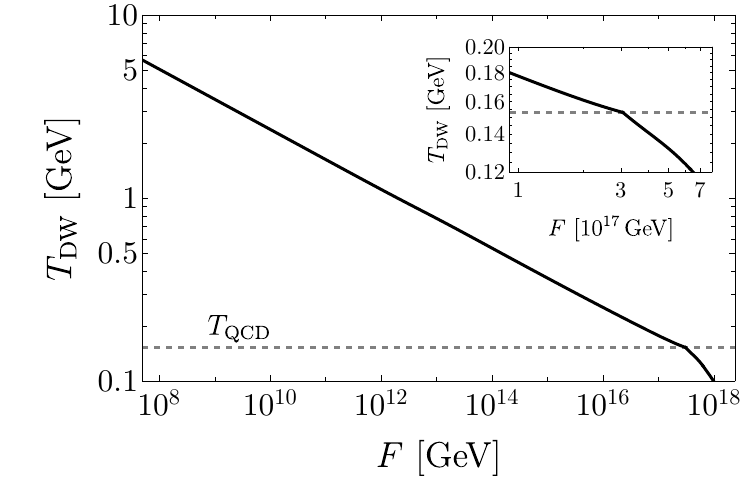}
    \caption{%
       $T_\mathrm{DW}$ as a function of $F$.
       The horizontal dashed line denotes $T_\mathrm{DW} = T_\mathrm{QCD}$.
       The behavior of $T_\mathrm{DW}$ changes on either side of the dashed line.
    }
    \label{fig: TDW}
\end{figure}

In the following, we provide a more detailed discussion of domain wall formation and the subsequent emergence of string bundles. 
The temperature of the domain wall formation is given by 
\begin{align}
    T 
    \simeq 
    T_\mathrm{DW} 
    \equiv 
    \begin{cases}
    \left( \dfrac{90 M_\mathrm{Pl}^2 \chi_0}{\pi^2 g_{*,\mathrm{DW}} F^2} \right)^{1/4}
    &
    (T_\mathrm{DW} < T_\mathrm{QCD})
    \\
    \left( \dfrac{90 M_\mathrm{Pl}^2 \chi_0 T_\mathrm{QCD}^n}{\pi^2 g_{*,\mathrm{DW}}F^2} \right)^{1/(n+4)}
    &
    (T_\mathrm{DW} \geq T_\mathrm{QCD})
    \end{cases}
    \ ,
    \label{eq: TDW}
\end{align}
where $M_\mathrm{Pl} \simeq 2.4 \times 10^{18}$\,GeV is the reduced Planck mass, and $g_{*,\mathrm{DW}}$ is the relativistic degree of freedom for the energy density at $T = T_\mathrm{DW}$.
We show the dependence of $T_\mathrm{DW}$ on $F$ in Fig.~\ref{fig: TDW}.
Here, we adopt the fitting function of $g_*(T)$ given in Ref.~\cite{Saikawa:2018rcs}.

\begin{figure}[t]
    \centering
    \includegraphics[width=.8\textwidth]{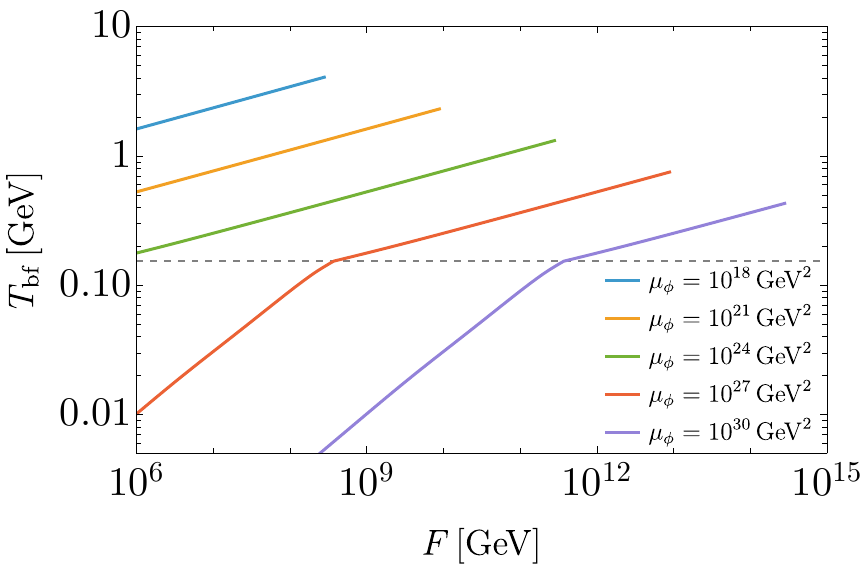}
    \caption{%
       The temperature when the string bundles form, $T_\mathrm{bf}$,  as a function of $F$.
       The colored lines represent different values of the tension of $\phi$ strings, $\mu_\phi$.
       The horizontal dashed line denotes $T_\mathrm{bf} = T_\mathrm{QCD}$.
       Here, we evaluate $T_\mathrm{bf}$ by taking the smaller value of $T_\mathrm{DW}$ in Eq.~\eqref{eq: TDW} and $T_\mathrm{bf}$ in Eq.~\eqref{eq: Tbf}.
    }
    \label{fig: Tbf}
\end{figure}

The formed domain walls have a tension
\begin{align}
    \sigma_A(T)
    \simeq
    8 m_A(T) F^2
    \ .
\end{align}
To form string bundles, the tension force of the domain walls must compete with that of the strings, so that they can pull the $\phi$-strings toward the $a$-strings. We expect that the domain wall tension force overcomes the string tension force when the energy density of domain walls exceeds that of strings. 
For simplicity, we focus on the case of $f_a/N_{\rm DW} \ll  f_\phi$ so that the QCD axion mostly consists of $a$.
In this case, the decay constant for the heavy axion (i.e., the QCD axion) is approximately given by $F \simeq f_a/N_{\rm DW}$.\footnote{
In the opposite case of $f_a/N_{\rm DW} \gg f_\phi$, the QCD axion mainly consists of $\phi$. For a given $f_a$, the string bundles start to form right after the domain walls appear, and the QCD axion abundance is considered to be suppressed compared to the case studied in the main text.
Note that, in this case,  there is no need to solve the domain wall problem from the beginning because $\phi$ has a unit domain wall number, but by adding an additional axion ($a$) one can similarly generate string bundles, which have the same interesting cosmological implications.
}
Assuming that both the string and wall networks are in the scaling regime, this domain-wall dominance condition becomes
\begin{align}
    \sigma_A(T) H \gtrsim \mu_\phi H^2
    \label{eq: bundle formation condition}
    \ ,
\end{align}
where $H$ is the Hubble parameter, and $\mu_\phi (\sim \pi f_\phi^2) $ is the tension of $\phi$-strings.
We can estimate the temperature at which the bundles start to form by solving Eq.~(\ref{eq: bundle formation condition}),
\begin{align}
    T 
    \simeq 
    T_\mathrm{bf}
    \equiv
    \begin{cases}
    \left( 
        \dfrac{2^7 \times 3^2 \times 5 }{\pi^2g_{*,\mathrm{bf}}} 
        \dfrac{M_\mathrm{Pl}^2 \chi_0 F^2}{\mu_\phi^2}
    \right)^{1/4}
    &
    (T_\mathrm{bf} < T_\mathrm{QCD})
    \\
    \left( 
        \dfrac{2^7 \times 3^2 \times 5 }{\pi^2g_{*,\mathrm{bf}}} 
        \dfrac{M_\mathrm{Pl}^2 \chi_0 F^2  T_\mathrm{QCD}^n}{\mu_\phi^2}
    \right)^{1/(n+4)}
    &
    (T_\mathrm{bf} \geq T_\mathrm{QCD})
    \end{cases}
    \ .
    \label{eq: Tbf}
\end{align}
We show the dependence of $T_\mathrm{bf}$ as a function of $F$ for $F < f_\phi/2$ with $\mu_\phi = \pi f_\phi^2$ in Fig.~\ref{fig: Tbf}.
In this region, $T_\mathrm{bf}$ is always smaller than $T_\mathrm{DW}$.
For values of the decay constant in the range $10^9\,\mathrm{GeV} \lesssim F \lesssim 10^{12}\,\mathrm{GeV}$, where the QCD axion does not overclose the Universe and still satisfies astrophysical bounds, the string bundles are found to start forming at temperatures around $\mathcal{O}(0.01$--$1)\,\mathrm{GeV}$ depending on $\mu_\phi$.

The typical size of a string bundle is given by ${\cal O}(1-10) \mu_\phi/\sigma_A(T)$, which is of order the Hubble radius at the bundle formation. As the Universe cools, both $\sigma_A(T)$ and $1/H$ increase, causing the bundle size to become smaller than the Hubble radius at later times. The typical size of the string bundle
is generally larger than the domain wall width $\sim 1/m_A$.
We also obtain the same result by considering the repulsion force between the $\phi$-strings.%
\footnote{Note that when the separation exceeds $1/m_A$, the strings interact only through the exchange of the massless axion.}
This analytical expectation is confirmed by our two-dimensional lattice simulations. See Appendix~\ref{appendix B} for details.

Let us estimate the QCD axion abundance in our scenario. 
Since the string bundles are formed some time after the domain wall formation, QCD axions continue to be produced from strings and domain walls until the string bundle formation begins. After the formation of the string bundles begins, however, QCD axion production is expected to be kinematically suppressed. This is because the condition
\begin{align}
    m_{A}(T_\mathrm{bf}) \gtrsim \frac{\sigma_A(T_\mathrm{bf})}{\mu_\phi}
\end{align}
is satisfied in this case. One can see this by noting that $\sigma_A = 8 m_A F^2$, $\mu_\phi \sim \pi f_\phi^2$, and $F < f_\phi/2$ (see Appendix~\ref{appendix} for more details). Note that light axions continue to be emitted from the string bundles. Therefore, the dominant contribution to the QCD axion abundance arises from the string-wall network just before the string bundle formation begins.

\begin{figure}[t]
    \centering
    \includegraphics[width=.8\textwidth]{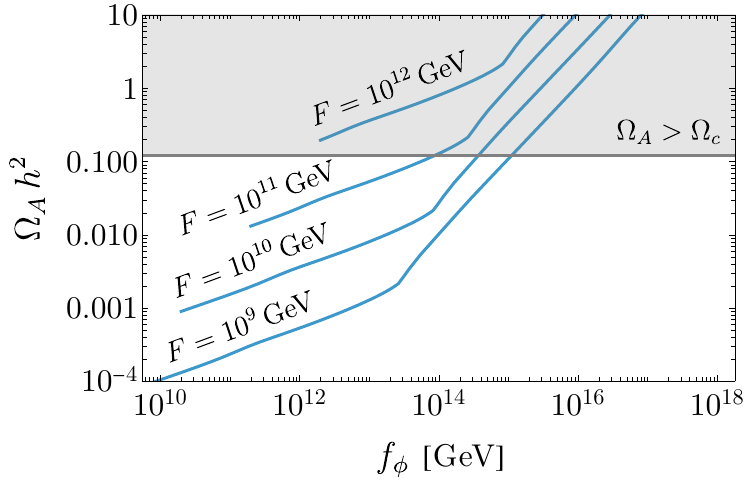}
    \caption{%
    The QCD axion abundance from the string-wall network as a function of the decay constant $f_\phi$, for the QCD axion decay constant $F = 10^{12},\,10^{11},\,10^{10}\,10^9$\,GeV from top to bottom. The gray shaded region indicates where the abundance exceeds the observed dark matter abundance $\Omega_c h^2 \simeq 0.12$.
    }
    \label{fig: Omega_A}
\end{figure}

Here, we follow the evaluation for the domain wall decay in the scaling regime~\cite{Kawasaki:2014sqa}.
In the scaling regime, the comoving area density of domain walls decreases, and their energy is emitted in the form of axion particles. 
We assume that the emitted axion has an energy of $\epsilon m_A$ with a constant parameter $\epsilon = \mathcal{O}(1)$.
Since axion particles are diluted faster than domain walls in the scaling regime, we can neglect the contribution of axions emitted well before the beginning of the bundle formation.
Then, the axion number density after the bundle formation is given by 
\begin{align}
    n_{A}
    \simeq
    \frac{\sigma_A(T_\mathrm{bf}) H_\mathrm{bf}}{\epsilon m_A(T_\mathrm{bf})} \left( \frac{R_\mathrm{bf}}{R} \right)^3
    \ ,
\end{align}
where $R$ is the scale factor, and the subscript `bf' denotes quantities at $T = T_{\rm bf}$.
Then, the energy density of the QCD axion in the later Universe is evaluated as 
\begin{align}
    \frac{\rho_A}{s_\mathrm{tot}}
    &=
    \frac{\sigma_A(T_\mathrm{bf}) H_\mathrm{bf}}{\epsilon \frac{2 \pi^2}{45}g_{*s,\mathrm{bf}} T_\mathrm{bf}^3}
    \frac{m_{A0}}{m_A(T_\mathrm{bf})}
    \nonumber \\
    &= 
    \begin{cases}
        \dfrac{3^{1/2} \times 5^{1/4}}{2^{1/4} \sqrt{\pi}}
        \dfrac{g_{*,\mathrm{bf}}^{3/4}}{\epsilon g_{*s,\mathrm{bf}}}
        \dfrac{\chi_0^{1/4}\sqrt{F \mu_\phi}}{M_\mathrm{Pl}^{3/2}} 
        &
        (T_\mathrm{bf} < T_\mathrm{QCD})
        \\
        \dfrac{1}{\epsilon g_{*s,\mathrm{bf}}}
        \left( 
            \dfrac{2^{3n-2} \times 45^{n+2} g_{*,\mathrm{bf}}^{n+6}}{\pi^{2n+4} } 
            \dfrac{\mu_\phi^4 \chi_0^{n+2} F^{2n+4}}{M_\mathrm{Pl}^{2n+12} T_\mathrm{QCD}^{2n}}
        \right)^{\frac{1}{2n+8}} 
        &
        (T_\mathrm{bf} \geq T_\mathrm{QCD})
    \end{cases}
    \ .
\end{align}
We show the abundance of the QCD axion produced by (partial) domain wall annihilation in Fig.~\ref{fig: Omega_A} for $F = 10^9,\,10^{10},\,10^{11},\,10^{12}$\,GeV. Here we set $\mu_\phi = \pi f_\phi^2$. One can see that, considering that the misalignment contribution becomes dominant for $F \gtrsim {\cal O}(10^{11})$\,GeV, the preferred value of $f_\phi$ lies around $10^{13\text{--}15}$\,GeV.\footnote{
The PQ symmetry of the scalar $\Phi$ can be restored even with a large decay constant if the reheating temperature is sufficiently high; it can also be restored at a lower reheating temperature or inflation scale when the $\Phi$ potential is flat, its nonminimal coupling to gravity is large, or $\Phi$ serves as the waterfall field in hybrid inflation.}
 This suggests that such string bundles may be probed by future CMB observations, pulsar timing arrays such as SKA, and gravitational experiments such as DECIGO, BBO, and LISA (see Ref.~\cite{Chang:2021afa}).
Note that, if the string tension grows logarithmically, the associated CMB and gravitational-wave signals are enhanced.
At large scales, string bundles can be approximated as ordinary global strings with an effective tension $N_{\rm DW}^2 \mu_\phi$.\footnote{This will be replaced by $\mu_ a \sim \pi f_a^2$ if we impose the inverse hierarchy, $f_a \gg N_{\rm DW} f_\phi$, where $\phi$ mainly plays a role of the QCD axion.}
Our model therefore provides a concrete example of a global string that can act as a source of gravitational waves without accompanying domain walls.
Massless axions $A_L$ are produced during the evolution of string bundles, but their energy density never dominates the universe; thus, there is no relevant constraint from $N_{\rm eff}$.
On the other hand, the internal structure of the string bundles could make characteristic features on the gravitational wave signatures from the ordinary global string case.
For the case with $f_\phi = 10^{15}\,{\rm GeV}$, the typical width of the string bundle is $\sim 10^{6-7}\,{\rm m}$, when the QCD axion explains whole dark matter.
The relative motion of $\phi$-strings could produce gravitational waves with the frequency $\sim 10^{1-2}\,{\rm Hz}$. More detailed study is left for future work.

On the other hand, for $F < 10^9$\,GeV, already subject to severe astrophysical bounds, the string bundle formation must be delayed in order to yield the right amount of the QCD axion dark matter.
This requires a much larger $f_\phi$, 
which in turn implies a large string tension $\mu_\phi$ that is likely already excluded by current observational constraints.
This issue can be circumvented if the string bundles subsequently annihilate, which can be achieved by introducing an additional potential as discussed below. 

So far, we have assumed that the axion potential arises solely from non-perturbative QCD effects. However, it is widely believed that continuous global symmetries are explicitly broken by quantum gravity effects, implying that there is no exactly massless axion. Therefore, let us comment on the case where an additional potential is present:
\begin{align}
    V_\Lambda(a,\phi)
    =
    \Lambda^4 \left[ 1 - \cos \left( n_a \frac{a}{f_a} + n_\phi \frac{\phi}{f_\phi} \right) \right]
    \ ,
\end{align}
where $n_a$ and $n_\phi$ are integers.
Here we assume that this potential is much smaller than $V_\mathrm{QCD}$ and does not affect the formation of string bundles.
Then, the massless mode generically acquires a mass due to $V_\Lambda$, and domain walls are subsequently formed around each string bundle.
The number of domain walls attached to a single string bundle is given by $N_\mathrm{DW}' = |n_a - N_{\rm DW} n_\phi|$ as shown in Ref.~\cite{Lee:2024toz}. 
If $N_\mathrm{DW}' \neq 0$, we have to consider the domain wall problem again.
This situation arises when the lighter axion acquires a mass from this potential. Thus, the extra potential can potentially revive the domain wall problem.
There are several ways to avoid the domain wall problem due to $V_\Lambda$: 
(i) the network of the string bundles and domain walls rapidly decays due to $N_\mathrm{DW}' = 1$ (e.g. $n_a = N_{\rm DW}+1, n_\phi = 1$), 
(ii) the tension of domain walls due to $V_\Lambda$ is smaller than $(1\,\mathrm{MeV})^3$,
(iii) the mass of the light mode is so small that domain walls are not yet formed by the current time.
Note that $(n_{a}, n_\phi) = (0, 1)$ does not solve the domain wall problem, although it would not induce the domain wall problem in the scenarios with no defects associated with $a$.

So far, we have focused on the case of $N_{\rm DW} \ne 1$ as our primary motivation is to solve the domain wall problem of the QCD axion. However, it is worth noting that string bundles can also form in the case of $N_{\rm DW} = 1$, provided that there is a light axion $\phi$ coupled to gluons. 

Let us now briefly comment on the LS mechanism~\cite{Lazarides:1982tw}.
In its canonical form, the model contains a gauged continuous symmetry. The color anomaly breaks the $U(1)_{\rm PQ}$ symmetry down to its subgroup $\mathbb Z_{N_{\rm DW}}$, which can be embedded in the center of the unbroken gauge group. As a result, the $N_{\rm DW}$ vacua are related by a gauge transformation and are therefore physically identical. That is to say, by introducing the gauge symmetry, the topology of vacua is modified such that there is no domain wall problem.\footnote{ See also Ref.~\cite{Lu:2023ayc}, which points out that even in models where the $\mathbb Z_{N_{\rm DW}}$ subgroup of $U(1)_{\rm PQ}$ is embedded in a continuous gauge group, the cosmological dynamics of string formation may still fail to eliminate domain walls, depending on the UV completion.} 
In terms of topological defects, each axionic string is then attached to only a single domain wall, so the string–wall network is topologically unstable and annihilates quickly. In fact, a similar dynamics is possible without embedding the $\mathbb{Z}_{N_{\rm DW}}$ into center symmetry~\cite{Sato:2018nqy}, in which case the axionic strings split into Alice strings. 
It is also possible to make use of a global symmetry, in which case the $N_{\rm DW}$ vacua are connected by a flat direction, i.e., a light or nearly massless NG mode~\cite{Barr:1982bb,Kawasaki:2015ofa}.
Our scenario has a similarity to this global-type LS mechanism in the sense that the $N_{\rm DW}$ vacua are connected by an additional light axion. Thus, the $N_{\rm DW}$ vacua correspond to the same physical vacuum plus a flat direction, so they can, in principle, be connected once an appropriate boundary of walls appears.
Whether such a boundary forms depends on the domain-wall number carried by the $\phi$–string: if this number is unity, the boundary is readily produced and the network reorganizes into stable string bundles; if it is two or larger, the wall network forms before the boundary can develop, and the domain wall problem cannot be solved~\cite{Lee:2024toz}.%
\footnote{Note that the domain wall problem is mitigated in this case, since the tension can be much smaller than the QCD axion domain wall.}
Unlike the LS mechanism, where all topological defects disappear, our scenario leaves behind stable string bundles as the final configuration. This possibility was discussed in Ref.~\cite{Takahashi:2020tqv} in the context of the cosmic birefringence, but the detailed study in the context of resolving the QCD axion domain wall problem and the axion dark matter production has not yet been carried out.

Our solution to the domain wall problem predicts the existence of a string-bundle network in the late-time Universe.
This remnant string bundle serves as a smoking-gun signature of our scenario.
If the massless axion couples to photons—as is often the case in UV completions\footnote{
This is indeed the case when $\phi$ couples to gluons as in the KSVZ axion model~\cite{Kim:1979if,Shifman:1979if}.
}—the stable string bundle induces anisotropic cosmic birefringence~\cite{Agrawal:2019lkr,Jain:2022jrp}.
Therefore, the simultaneous discovery of the QCD axion dark matter and anisotropic cosmic birefringence, as well as stochastic gravitational-wave backgrounds across multiple frequency bands-from the nanohertz range accessible to pulsar timing arrays to the millihertz–hertz range probed by space- and ground-based interferometers-would provide compelling evidence for this scenario.
Alternatively, the light axion may be probed via future observations of gamma rays or X-rays through axion–photon conversion, particularly if the decay constant is smaller than $10^{10\text{--}11}\,\mathrm{GeV}$.
Intriguingly, this mechanism might even explain certain observed events~\cite{Nakagawa:2022wwm}.

\section{Solution with transient bias}
\label{sec: transient}

Here, we discuss another possibility for solving the domain wall problem: the collapse of domain walls due to a transient bias.
In this section, we consider the case where $\Phi$ has a pre-inflationary initial condition, so that $\phi$ does not form strings.
However, before $\phi$ starts oscillating around its potential minimum, its potential acts as an effective bias term, destabilizing the string-wall network and leading to the collapse of domain walls.
Since this bias disappears once $\phi$ begins oscillating, the strong CP phase is not shifted in the vacuum, and the PQ mechanism remains intact.

\subsection{Setup}
\label{subsec: transient setup}

We again introduce another axion, $\phi$, but it mixes with $a$ in an additional potential term:
\begin{align}
    V(a,\phi)
    &= V_\mathrm{QCD}(a) 
     + V_\Lambda(a,\phi) 
    \nonumber \\
    &=
    \chi(T) \left[ 1 - \cos \left( N_{\rm DW} \frac{a}{f_a} \right) \right]
    + \Lambda^4 \left[ 1 - \cos \left( n_a \frac{a}{f_a} + n_\phi \frac{\phi}{f_\phi} \right) \right]
    \ ,
\end{align}
where $n_a$ and $n_\phi$ are integers.\footnote{
The mixing in the coupling to gluons, together with the potential of $\phi$, can be cast into this form by the unitary transformation~\cite{Murai:2024nsp}.
} 
Here, $V_\mathrm{QCD}$ and $V_\Lambda$ come from the non-perturbative effects of QCD and the hidden sector, respectively.
We set the origins of $a$ and $\phi$ so that $a = \phi = 0$ corresponds to the potential minimum.
For later convenience, we define
\begin{align}
    m_{a,\Lambda}
    \equiv 
    \frac{n_a \Lambda^2}{f_a}
    \ .
\end{align}

Here, we assume the formation of a string-wall network associated with the axion $a$, while the axion $\phi$ is taken to have a homogeneous initial condition.
We further assume a hierarchy of mass scales and decay constants
\begin{align}
    m_{a,\Lambda} \ll m_{a0}
    \ , \quad 
    \epsilon \equiv \frac{n_\phi}{n_a} \frac{f_a}{f_\phi} \ll 1
    \ ,
    \label{eq: hierarchy for transient}
\end{align} 
which ensures that the mass eigenstates are primarily determined by $V_\mathrm{QCD}$ at low temperatures.

Before discussing the evolution of the defects, we examine the mass eigenstates at the potential minimum.
Around the minimum at $a = \phi = 0$, the potential can be approximated by  quadratic terms as
\begin{align}
    V(a,\phi)
    &\simeq 
    \frac{m_a^2}{2} a^2
    +
    \frac{\Lambda^4}{2 f_a^2} \left( n_a a + \frac{n_\phi f_a}{f_\phi} \phi \right)^2
    \nonumber \\
    &=
    \frac{1}{2} 
    \begin{pmatrix}
        a & \phi
    \end{pmatrix}
    \mathcal{M}^2
    \begin{pmatrix}
        a \\ \phi
    \end{pmatrix}
    \ ,
\end{align}
with 
\begin{align}
    \mathcal{M}^2
    =
    \begin{pmatrix}
        m_a^2 + \dfrac{n_a^2 \Lambda^4}{f_a^2} 
        &
        \dfrac{n_a n_\phi \Lambda^4}{f_a f_\phi}
        \vspace{2mm}
        \\
        \dfrac{n_a n_\phi \Lambda^4}{f_a f_\phi} 
        &
        \dfrac{n_\phi^2 \Lambda^4}{f_\phi^2}
    \end{pmatrix}
    =
    \begin{pmatrix}
        m_a^2 + m_{a,\Lambda}^2
        &
        \epsilon m_{a,\Lambda}^2
        \vspace{2mm}
        \\
        \epsilon m_{a,\Lambda}^2
        &
        \epsilon^2 m_{a,\Lambda}^2
    \end{pmatrix}
    \ .
\end{align}
The mass matrix can be diagonalized as
\begin{align}
    M_d^2 
    \equiv 
    \begin{pmatrix}
        m_H^2
        &
        0
        \vspace{2mm}
        \\
        0
        &
        m_L^2
    \end{pmatrix}
    \equiv
    \mathcal{O} \mathcal{M}^2 \mathcal{O}^T
    \ ,
\end{align}
where
\begin{gather}
    m_{H,L}^2 
    =
    \frac{1}{2} \left[
        m_a^2 
        + ( 1 + \epsilon^2 ) m_{a,\Lambda}^2 
        \pm \sqrt{ 
            m_a^4 
            + 2 ( 1 - \epsilon^2 ) m_a^2 m_{a,\Lambda}^2 
            + ( 1 + \epsilon^2 )^2 m_{a,\Lambda}^4 
        }
    \right]
    \ ,
    \\
    \mathcal{O}
    =
    \begin{pmatrix}
        \cos \alpha
        &
        -\sin \alpha
        \\
        \sin \alpha
        &
        \phantom{-} \cos \alpha
    \end{pmatrix}
    \ , \quad 
    \tan \alpha 
    =
    \frac{m_a^2 + (1 - \epsilon^2) m_{a,\Lambda}^2 
    - \sqrt{
        m_a^4 
        + 2 ( 1 - \epsilon^2 ) m_a^2 m_{a,\Lambda}^2 
        + ( 1 + \epsilon^2 )^2 m_{a,\Lambda}^4 
    }}{2 \epsilon m_{a,\Lambda}^2}
    \ .
\end{gather}
Up to the leading order of $\epsilon$, we obtain
\begin{gather}
    m_H^2 
    \simeq 
    m_a^2 + m_{a,\Lambda}^2 
    \ , \quad
    m_L^2 
    \simeq 
    \epsilon^2 \frac{m_a^2 m_{a,\Lambda}^2}{m_a^2 + m_{a,\Lambda}^2}
    \ ,
    \\
    \tan \alpha 
    \simeq
    - \epsilon \frac{ m_{a,\Lambda}^2}{m_a^2 + m_{a,\Lambda}^2}
    \ .
\end{gather}

The mass eigenstates are
\begin{align}
    \begin{pmatrix}
        \phi_H
        \\
        \phi_L
    \end{pmatrix} 
    =
    \mathcal{O}
    \begin{pmatrix}
        a \\ \phi
    \end{pmatrix}
    \simeq 
    \begin{pmatrix}
        a + \epsilon \frac{ m_{a,\Lambda}^2}{m_a^2 + m_{a,\Lambda}^2} \phi
        \\
        \phi - \epsilon \frac{ m_{a,\Lambda}^2}{m_a^2 + m_{a,\Lambda}^2} a
    \end{pmatrix}
    \ .
\end{align}
Thus, the axion oscillation can be separated into $a$ and $\phi$ for $\epsilon \ll 1$.

\subsection{Dynamics of the string-wall network}
\label{subsec: evolution}

As the cosmic temperature drops to the QCD scale, $\chi$ increases, and the axion field $a$ begins to oscillate around the minimum of $V_\mathrm{QCD}$. Once domain walls are formed, $a$ settles into the potential minimum away from the walls, while $\phi$ remains at its initial value. 
We assume that $\phi$ has a spatially homogeneous initial value throughout the Universe.

Once $\phi$ begins to evolve due to $V_\Lambda$, both $a$ and $\phi$ roll down to the degenerate minima within each domain, and no potential bias remains.
In contrast, while $\phi$ is still frozen at its initial value due to its small mass, the potential energy generally varies across different domains. This energy difference acts as a (transient) potential bias, which can induce the collapse of the string-wall network.
In this sense, the setup effectively introduces a transient explicit breaking of the PQ symmetry in the early Universe, which disappears later due to the dynamics of the lighter ALP~\cite{Lee:2024xjb}.

To be specific, we assume
\begin{align}
    N_{\rm DW} = 2, 
    \quad 
    n_a = 1, 
    \quad 
    n_\phi = 1
    \ ,
\end{align}
in the following analysis, but our results can be straightforwardly applied to more general cases.
We further assume that the lighter degree of freedom remains frozen until well after the domain wall network collapses.%
\footnote{
In fact, this is a rather conservative condition for the collapse of domain walls. The transient bias could induce a population bias in the domain wall network, which results in a subsequent decay of the domain walls. The decay of domain walls due to population bias for various types of initial fluctuations was studied in Refs.~\cite{Gonzalez:2022mcx,Kitajima:2023kzu}.
}
The onset of $\phi$ oscillation is estimated by 
\begin{align}
    m_L \sim H_{\rm osc}
    \ .
\end{align}
So we require
\begin{align}
    H_{\rm osc} \lesssim H_{\rm dec},
    \label{cond}
\end{align}
where $H_{\rm dec}$ denotes the Hubble parameter at the time of domain wall decay. Hereafter, the subscript ``dec'' denotes the quantity evaluated at the epoch of domain wall network decay.

With $V_\Lambda$ neglected, domain walls whose tension is determined by $V_\mathrm{QCD}$ separate domains where $a \simeq 0$ or $\pi f_a$.
The domain wall tension is given by
\begin{align}
    \sigma = \frac{8 m_H f_a^2}{N_\mathrm{DW}^2} = 2m_H f_a^2
    .
\end{align}
If $\phi$ remains at its initial value, the potential difference due to $V_\Lambda$ is given by
\begin{align}
    \Delta V = 2 m_{a,\Lambda}^2 f_a^2 
    \left| \cos \left( \frac{\phi_{\mathrm{ini}}}{f_\phi} \right) \right|
    \equiv 2\, c\, m_{a,\Lambda}^2 f_a^2 \ . 
    \label{bias}
\end{align}

The condition for the domain wall network collapse is 
\begin{align}
    \sigma H_{\rm dec} \sim \Delta V \ .
    \label{collapse}
\end{align}
Eq.~\eqref{cond} then leads to%
\footnote{Note that when this condition is satisfied, the induced domain walls that would otherwise stabilize the network, as pointed out in Refs.~\cite{Lee:2024xjb,Lee:2024toz}, do not form.}:
\begin{align}
\label{eq:condition on tension}
    m_L \lesssim \frac{\Delta V}{\sigma}
    \quad \Leftrightarrow \quad
    \epsilon \lesssim
    c \frac{m_{a,\Lambda}}{m_H}
    \ .
\end{align}
Therefore, the string-wall network decays before the transient bias disappears, if $\epsilon$ is sufficiently small. 
We then obtain
\begin{align}
    m_H \lesssim c  \frac{f_\phi^2}{f_a^2} m_L \ ,
\end{align}
where we have used $m_L \simeq \Lambda^2 / f_\phi$ which follows from $m_{a,\Lambda} \ll m_a$.
This inequality, together with the assumption $m_a > H_{\rm dec} > m_L$, can be satisfied when $f_\phi \gg f_a$.

Next, let us consider the possibility that QCD axions emitted during the domain wall collapse constitute the dominant component of dark matter. Assuming that an $\mathcal{O}(1)$ fraction of the domain wall energy is transferred to axion dark matter, we obtain:
\begin{align}
    \frac{\rho_A}{s_{\rm tot}} 
    \sim 
    \frac{\sigma H_{\rm dec}}{m_a(T{_{\rm dec}}) s_{{\rm tot},\,{\rm dec}}} m_{a,0} 
    \sim 
     10^{-18}\,{\rm GeV}^{-1} 
    \frac{f_a^2 m_{a,0}}{g^{1/2}_{*,
    \rm dec}  T_{\rm dec}}  \ ,
    \label{eq: transient QCD axion abundance}
\end{align}
where the temperature dependent $\chi$ cancels out.

Requiring that this accounts for the observed dark matter abundance gives
\begin{align}
    \label{axionDM}
    f_a \sim 2\times 10^8\,{\rm GeV}
    \left(\frac{g_{*,\rm dec}}{10.6}\right)^{1/2} \left( \frac{T_{\rm dec}}{1\,{\rm MeV}} \right) \ .
\end{align}
For the QCD axion with a relatively low decay constant ($f_a < 10^{10}$ GeV) to constitute the dominant dark matter, we must have $T_{\rm dec} < T_{\rm QCD}$.
For simplicity, we do not consider a lighter QCD axion ($f_a > 10^{10}$ GeV), where both the string network and misalignment contributions should be taken into account.

The upper and lower bounds of the ALP decay constant, $f_{\phi ,\, \rm max}$ and $f_{\phi ,\, \rm min}$, required for successful QCD axion dark matter production are determined by avoiding the overproduction of the light ALP and and by requiring that the domain walls decay due to the transient bias, Eq.~(\ref{cond}).
At the time of collapse, the axion mass is given approximately by its vacuum value:
\begin{align}
m_a(T{_{\rm dec}}) \approx m_{a0} = 2\times \frac{(75.6\,{\rm MeV})^2}{f_a} ,
\end{align}
{and then $\Lambda$ is related to the Hubble parameter at the domain wall network collapse through the condition~(\ref{collapse}) as
\begin{align}
    \Lambda \sim 0.03\,{\rm MeV} 
    \left( \frac{f_a}{10^8\,{\rm GeV}} \right)^{1/4} 
    \left( \frac{g_{*,\,\rm dec}}{10.6} \right)^{1/8} 
    \left( \frac{T_{\rm dec}}{{\rm MeV}} \right)^{1/2} 
    c^{-1/4} \ .
\end{align}
The condition for the transient bias to remain until the domain wall decay, $m_L < H_{\rm dec}$, requires
\begin{align}
    f_\phi > f_{\phi,\,\rm min} \sim 2 \times 10^{15}\,{\rm GeV} 
    \left( \frac{1\,{\rm MeV}}{T_{\rm dec}} \right) 
    \left( \frac{10.6}{g_{*,\,\rm dec}} \right)^{1/4} 
    \left( \frac{f_a}{10^8\,{\rm GeV}} \right)^{1/2}
    c^{-1/2}\ .
\end{align}
The requirement that the light ALP not be overproduced sets the upper limit $f_{\phi,,\rm max}$.
The misalignment mechanism produces the lighter ALP, with an abundance given by
\begin{align}
    \Omega_L h^2
    \sim 0.03\, g_{*,\,\rm osc}^{-1/4}\, 
    \frac{\Lambda}{0.03\,{\rm MeV}} 
    \left( \frac{\phi_{\rm ini}}{f_\phi} \right)^{2} \left(\frac{f_\phi}{10^{15} \rm GeV}\right)^{3/2} 
    \ .
\end{align}
Avoiding the overproduction for a typical initial condition $\phi_\mathrm{ini} \sim f_\phi$ imposes an upper bound on $f_\phi$, which corresponds to the largest value of $f_\phi$ that yields $\Omega_L h^2 = 0.12$. 
Fig.~\ref{fig: AL_constraint} shows the parameter region where the above scenario works.
The blue shaded region corresponds to the overproduction of the light ALP with $f_a = 2 \times 10^8\,{\rm GeV}$ for the initial condition $\phi_{\rm ini} = f_\phi$. 
The orange shaded region indicates where the light ALP $A_L$ settles to the potential minimum before the domain wall decay.
The blue and orange dashed lines show the corresponding cases for   $f_a = 2 \times 10^9\,{\rm GeV}$.
Note that the lower bound on $f_\phi$ is a conservative one, since the population bias remaining even after the transient bias is turned off could still drive the domain walls to collapse. With a mild tuning of $\phi_{\rm ini}/f_\phi$, one can also suppress $\Omega_L$ while keeping $m_L$ sufficiently small for the transient bias to be effective. In fact, even when $\phi_{\rm ini} = 0$, the transient bias remains nonzero (with $c=1$), as can be seen in Eq.~\eqref{bias}. With these caveats in mind, we can see from this figure that the parameter region where the QCD axion constitutes the majority of dark matter is rather small, being limited to $f_\phi \approx (2 \text{\,--\,} 3) \times 10^{15}$\,GeV and $T_{\rm dec} \approx 1$\,MeV (see Eq.~(\ref{axionDM})). For higher $T_{\rm dec}$ along the blue solid line, the ALP is the dominant dark matter while the QCD axion is subdominant. The situation does not change for a larger $f_a$. This result is expected due to the condition (\ref{eq:condition on tension}). In this scenario, the ALP abundance tends to exceed that of the QCD axion unless the initial displacement $\phi_{\rm ini}$ is fine-tuned.

Lastly, let us mention the isocurvature constraint on $\phi$. Since we assume the pre-inflationary scenario for $\phi$, it acquires quantum fluctuations during inflation, which result in the isocurvature perturbation of $\phi$. This places an upper limit on the inflationary Hubble parameter, but it can be relaxed compared to the one for the QCD axion dark matter scenario since the required decay constant $f_\phi$ is larger.

\begin{figure}[t]
    \centering
    \includegraphics[width=.8\textwidth]{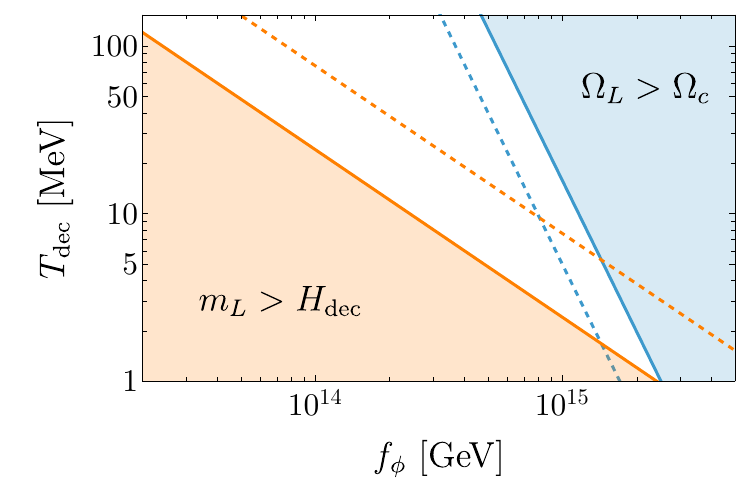}
    \caption{%
    Constraints in the $(f_\phi,T_{\rm dec})$ plane with $c=0.5$ and $\phi_{\rm ini}=f_\phi$.
    The blue-shaded region indicates overproduction of the light ALP, while the orange-shaded region shows where $A_L$ settles to its potential minimum before the domain-wall decay.
    The solid and dashed lines correspond to $f_a = 2\times10^8\,{\rm GeV}$ and $2\times10^9\,{\rm GeV}$, respectively.
    }
    \label{fig: AL_constraint}
\end{figure}

\section{Summary and discussion}
\label{sec: summary}

The PQ mechanism solves the strong CP problem and provides the QCD axion, which is a candidate for dark matter.
In the post-inflationary PQ breaking scenario, a string-wall network becomes long-lived for a domain wall number larger than unity, leading to the cosmological domain wall problem.
While a potential bias can destabilize the string-wall network, it reintroduces a fine-tuning of the minimum of the bias potential to solve the strong CP problem.

In this paper, we have proposed two simple solutions to the domain-wall problem induced by the QCD axion by introducing a second, massless (or extremely light) axion:
\begin{enumerate}
  \item \textbf{String-bundle confinement}:  
        If the new axion’s PQ symmetry breaks \emph{after}
        inflation, the associated strings can be connected to those of the QCD axion by domain walls.
        The domain walls then terminate on (and are confined
        by) these strings, forming cosmologically benign ``string bundles,'' instead of extended domain walls. 
        These bundles behave as cosmic strings and are much less problematic than domain walls, thereby solving the domain wall problem.
        Such string bundles can survive to the present epoch and generate cosmic
        birefringence. If the decay constant $f_\phi$ is sufficiently large, the string bundles may also leave observable imprints on the CMB anisotropies {as well as stochastic gravitational waves}, providing an additional avenue for detection. A simultaneous detection of the QCD axion and a
        birefringence signal would be a striking experimental signature.

  \item \textbf{{Transient} bias mechanism}:  
        If the new axion starts with a homogeneous initial value and is
        sufficiently light, its potential temporarily biases the QCD-axion
        potential \emph{before} the light axion begins to oscillate. 
        This induced bias destabilizes and eliminates the walls while leaving the PQ solution to the strong-CP problem intact. 
        In this case, both the QCD axion and ALP can contribute to dark matter depending on the model parameters.
\end{enumerate}
Thus, a light auxiliary axion that mixes with the QCD axion provides phenomenologically consistent, experimentally testable resolutions of the QCD-axion domain-wall problem without reintroducing the strong-CP fine-tuning.

Finally, we make brief comments on possible extensions of our scenarios.
If we consider more axions than two in the string bundle scenario, the formation condition for the bundle is expected to be more complicated.
Even if one axion has a domain wall number of unity, it does not guarantee bundle formation with large domain wall numbers for other axions or many axions with domain wall numbers larger than unity.
In addition, with three or more axions, the combination of strings and anti-strings to form bundles is no longer unique.

Related to this, we briefly comment on the connection with the clockwork mechanism.
The clockwork mechanism was first applied to the QCD axion in Ref.~\cite{Higaki:2015jag}.
In the clockwork QCD axion model, by appropriately assigning charges to many axions, one can generate a large hierarchy between the actual symmetry-breaking scale and the decay constant of the QCD axion.
In the minimal setup, the QCD axion is the lightest among the axions, while the others are heavy.
Since the clockwork mechanism involves a large number of axions, the internal structure of the corresponding string bundle becomes complicated.
In particular, a hierarchical difference is required in the number of constituent axion strings forming the bundle.
As a result, such string bundles do not form in realistic cosmological evolution; instead, domain walls associated with the heavy axions are expected to appear~\cite{Higaki:2016jjh,Long:2018nsl}.
Even in a two-axion system, the formation of string bundles is not guaranteed, and their possible configurations have been classified in Ref.~\cite{Lee:2024toz}.
In this paper, based on such a two-axion framework, we consider a different setup in which the QCD axion is the heavy axion while the lighter one plays the role of an ALP, and we show that in this case the domain-wall problem of the QCD axion can be resolved through the formation of string bundles.
We also note that, as mentioned in the main text, a similar model and dynamics in which the QCD axion and the ALP strings form a string bundle have already been pointed out in the context of cosmic birefringence in Ref.~\cite{Takahashi:2020tqv}.

We can also consider the transient bias in a setup where both of the two axions have cosmic strings, and string bundles are not effectively formed due to the domain wall numbers larger than unity.
In this case, the light axion becomes highly inhomogeneous, and the bias potential is also inhomogeneous even in a single domain.
Thus, it is unclear whether the potential bias can still trigger the collapse of the network.
To clarify the dynamics of such systems, dedicated lattice simulations will be necessary, which is
left for future work.

\section*{Acknowledgments}
We thank Steven Cotterill, Jiji Fan, Mark Hindmarsh, Javier Redondo, and Ken'ichi Saikawa for useful comments at {\it Axions in Stockholm 2025} (23rd June to 11th July, 2025), where the solution using the string bundles was presented.
This work is supported by JSPS KAKENHI Grant Numbers 25H02165 (F.T.), 23KJ0088 (K.M.), 24K17039 (K.M.), 22K14029 (W.Y.), 22H01215 (W.Y.), and 25KJ0564 (J.L.), Selective Research Fund for Young Researchers from Tokyo Metropolitan University (W.Y.), Graduate Program on Physics for the Universe (J.L.), and JST SPRING Grant Number JPMJPS2114 (J.L.). 
This article is based upon work from COST Action COSMIC WISPers CA21106, supported by COST (European Cooperation in Science and Technology).

\appendix

\section{Typical size of string bundles}
\label{appendix B}

Here we present our two-dimensional numerical lattice simulations of string-bundle formation. We set $N_{\rm DW} = 2$ to ensure that the formation of string bundles occurs within the accessible simulation time.

Fig.~\ref{fig: string-wall snapshot} shows snapshots of the string-wall system at an early time (left) and a late time (right). One can see that the string-wall network collapses into string bundles at late times. In the case of $N_{\rm DW} = 2$, the string bundle resembles a CO$_2$ molecule in shape, 
with a linear configuration consisting of a central $a$-string and two $\phi$-anti-strings attached at both ends.
We obtained its size, i.e., the distance between two $\phi$-strings assuming a linear configuration, by measuring the domain wall length for each string bundle.
Its size is approximately given by
\begin{align}
d \simeq 2\times \frac{1}{m_{A0}} \frac{f_\phi^2}{F^2},
\end{align}
as we analytically expected in Sec.~\ref{sec: string bundle}.
In the simulation, we choose $f_\phi = f_a \simeq 7.1 m_{A0}$ and constant $m_A = m_{A0}$ to synchronize the string formations of both axions and reduce the computational cost.
\begin{figure}[t]
    \center{
        \includegraphics[width=.45\textwidth]{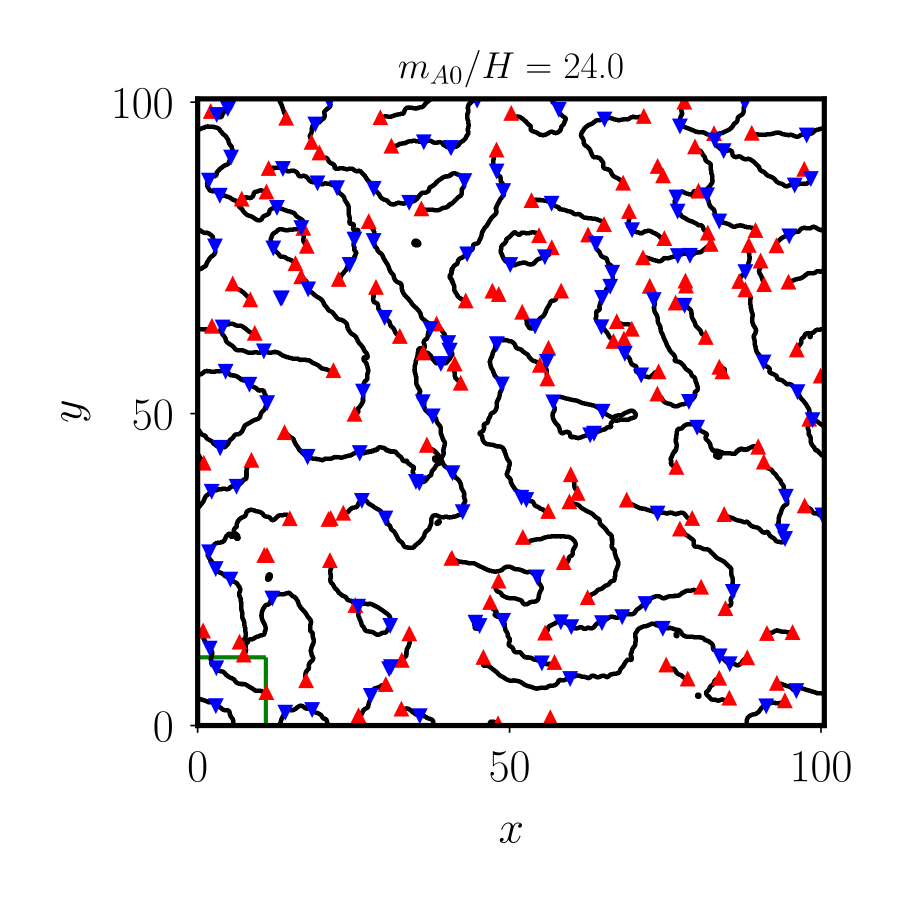}
        \includegraphics[width=.45\textwidth]{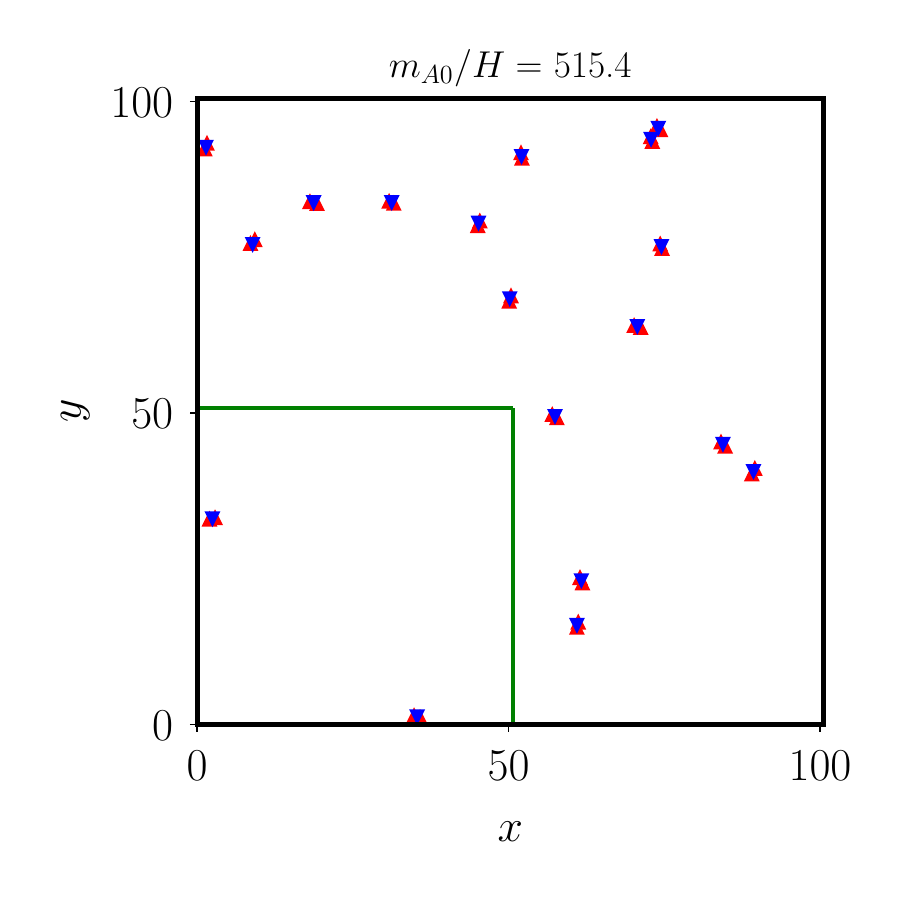}
    }
    \caption{%
        The snapshot of the string-wall system that appears in the $N_{\rm DW} = 2$ case at an early time (left) and a late time (right).
        The blue(red) triangles indicate the $a$-($\phi$-)string locations without distinction of the direction, and the black solid lines indicate the domain wall locations.
        For reference, we show the Hubble horizon scale using the green square.
        For simplicity, two kinds of axions have the same decay constant, and the mass is constant in the simulations.
        One can observe that the string bundles are formed in the late time snapshot.
    }
    \label{fig: string-wall snapshot}
\end{figure}

\section{Particle production from topological defects}
\label{appendix}

Let us consider a generic interaction Hamiltonian ${\cal H}_\mathrm{int}$ of fields $\phi_i=\bar\phi_i+\delta \phi_i$ including the background field $\bar\phi_i(t,\vec x)$. 
$i$ is the index of a field. 
The background field may depend on $t$ and $\vec x$. For instance, it may represent a collapsing domain wall configuration due to a potential bias, and alternatively it may be a configuration of the string bundle with fluctuations. 

To study the particle production, we consider using perturbation theory(see Ref.\,\cite{Vilenkin:2000jqa} for analogous discussion of radiation of scalar field from cosmic string). 
For simplicity, let us consider a single-particle production. The amplitude to produce $\phi_1$ is
\begin{align}
\langle{\phi_1(p)}| \hat T e^{i \int d^4x { \hat {\cal H}}_{\rm int}}|0\rangle \propto \int d^4x e^{i p x}(\partial_{\phi_1} {\cal H}_{\rm int })|_{\phi_i=\bar{\phi}_i(t,\vec{x})}
\end{align}
Here $p^0= \sqrt{\vec{p}^2+m_1^2}$ since we assume the asymptotic state for the perturbation theory. In the right-hand side, there is no ``hat'' for the Hamiltonian because it is only with the background field. 

To be more concrete, let us for simplicity consider some oscillating domain wall background with a period of $2\pi\mu_\phi/\sigma$, which is the case for the string bundle fluctuation. We take the vanishing limit of the width and express the domain wall as classical background. 
Then the target space coordinate of the domain wall is 
\begin{align}
x^\mu_\mathrm{DW}(\xi),
\end{align}
where $\xi$ is the world volume coordinates. 
Then we get 
\begin{align}
\langle{\phi_1(p)}| \hat T e^{i \int d^4x { \hat {\cal H}}_{\rm int}}|0&\rangle \propto \int d^4x d^3 \xi \sqrt{-\eta_\mathrm{DW}} e^{i p x} \delta^4(x-x_{\rm DW})\\
& \propto \int dt d^2 \xi_{(2)} \sqrt{-\eta_\mathrm{DW}} e^{iE t}  e^{- i \vec p \vec x_{\rm DW}(t,\xi_{(2)}) } 
\end{align}
with $\eta_{\rm DW}$ being the metric in the world volume coordinate. 
In the last row, we performed the $x$ integration and fixed the gauge with $\xi^0=t$ without loss of generality. The important point is that $\vec{x}_{\rm DW}$ is periodic in time with a period of $T=2\pi\mu_\phi/\sigma$. 
Thus 
\begin{align}
\langle{\phi_1(p)}| \hat T e^{i \int d^4x { \hat {\cal H}}_{\rm int}}|0&\rangle  
\propto \int dt \sqrt{-\eta_\mathrm{DW}} e^{iE t} \sum_{n}c_n e^{i\frac{2\pi n}{T} t }.
\end{align}
This argument implies that if $E> m_a \gg 2\pi/T = \sigma/\mu_{\phi},$ one needs to pick up higher harmonics, which are usually suppressed. 

Some loopholes to this argument are as follows. 
If the domain wall configuration has some cusps, one can get enhanced higher modes, and one can have significant particle production. 
For instance, for the collapsing domain walls, the domain wall would disappear, and the behavior is obviously singular. Thus the Fourier transformation of time allows high-energy modes. 

So far, we have focused on the domain wall, but for any more complex backgrounds that are oscillating with the period $T$, our discussion should hold, i.e., $\partial_{\phi_1} H_{\rm int}|_{\phi_i=\bar{\phi}_{i}}$ can be expressed by the harmonics controlled by $2\pi/T$. 
Then the production of higher-mass modes than $2\pi/T$ is suppressed. 
For multi-particle productions, the suppression is more efficient since the required energy is larger.
Although, if there is a cusp or collapse, the discussion may be avoided, it might be safer to concentrate on the region $\sigma/\mu_{\phi} > m_{a}$ for the QCD axion production. 

Fig.~\ref{fig: evol_phiH_density} shows the time evolution of the heavy axion energy density obtained in lattice simulations.
Here, to avoid contamination of the spectra due to the defects, we masked the strings and domain walls.
The basic setup is shared with the simulation in Fig.~\ref{fig: string-wall snapshot}, but we choose $f_\phi = 1.5 f_a \simeq 10.6 m_{A0}$ here.
The simulation implies that the final abundance is $2$\,--\,$3$ times larger than the rough analytic estimate shown as the red horizontal dashed line. Thus, our analytic estimate gives the correct axion abundance with an accuracy of ${\cal O}(1)$ factor.
We also confirmed that marginally non-relativistic heavy axions are produced at early times, and the comoving energy density freezes at late times.

Let us here comment on possible caveats of the numerical simulations.
Due to the radial oscillations of the complex scalar field and the masking method, the abundance exhibits oscillatory behavior at early times, making it difficult to identify when particle production terminates. Nevertheless, we can clearly see that the abundance freezes shortly after $H = H_\mathrm{bf}$.
Note however that the timing when the string bundles begin to form is relatively close to the onset of the axion oscillations due to computational limitations, and we need more dedicated simulations for quantitative estimates of the final axion abundance. 
If the axion abundance is enhanced, the decay constant $f_\phi$ required to account for the entire dark matter can be smaller, which in turn relaxes the constraints from gravitational signals.

\begin{figure}[t]
    \center{
        \includegraphics[width=.65\textwidth]{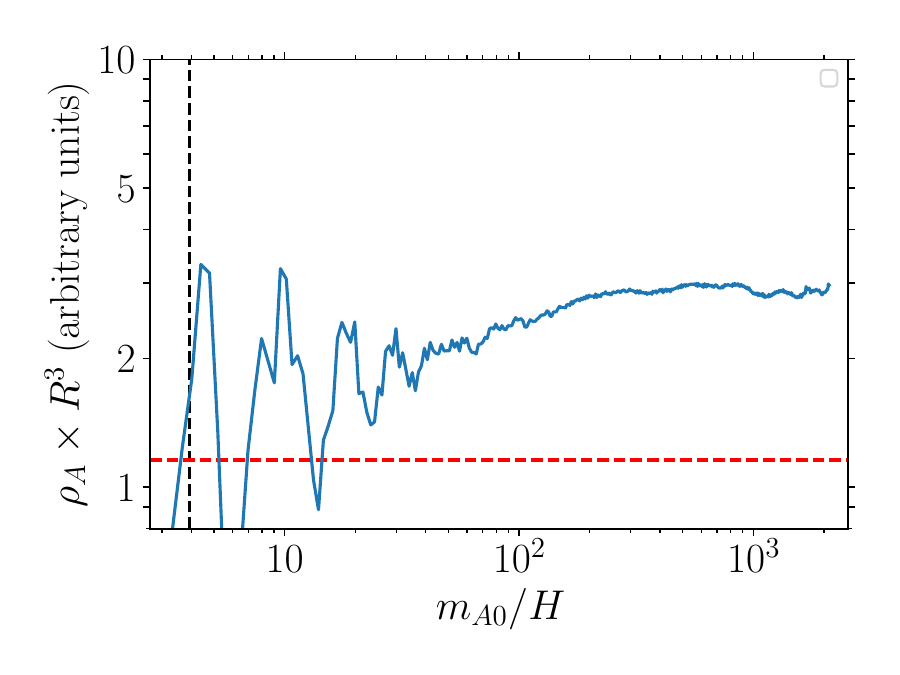}
    }
    \caption{%
        The time evolution of the heavy axion energy density multiplied by the scale factor cubed.
        The horizontal axis is the Hubble parameter inverse normalized by the heavy axion mass.
        The black vertical dashed line corresponds to $H = H_\mathrm{bf} = \sigma_A / \mu_\phi$ and the red horizontal dashed line corresponds to the expected abundance assuming instant termination of the particle production at $H = H_\mathrm{bf}$.
        We choose $f_\phi = 1.5 f_a \simeq 10.6 m_{A0}$.
        The heavy axion production is deactivated, and the yield freezes at a late time.
        The oscillations at the early times are considered to be due to the zero mode oscillations in the radial component of the complex scalar, which remain after the masking.
    }
    \label{fig: evol_phiH_density}
\end{figure}

\bibliographystyle{apsrev4-1}
\bibliography{ref}

\end{document}